\documentclass[useAMS]{mn2e}

\usepackage{graphicx}

\title[Orbital elements, masses and distance of $\lambda$\,Sco]{Orbital elements, masses and distance of
 $\lambda$ Scorpii A and B determined with the Sydney University Stellar Interferometer and
 high resolution spectroscopy}

\author[W. J. Tango et al.]
       {W. J. Tango,$^1$\thanks{E-mail: W.Tango@physics.usyd.edu.au}
        J. Davis,$^1$ M. J. Ireland,$^2$
        C. Aerts,$^{3,4}$  K. Uytterhoeven,$^{5,3}$
         \newauthor A. P. Jacob,$^1$ A. Mendez,$^1$ J. R. North,$^1$ E. B. Seneta,$^6$
          P.~G.~Tuthill$^1$\\
       $^1$Chatterton Astronomy Department, School of Physics, University of Sydney, NSW 2006,
       Australia\\
       $^2$Planetary Science, MS~150-21, Caltech, 1200 E. California Blvd, Pasadena CA 91125, USA\\
       $^3$Institute of Astronomy, Catholic University of Leuven, Celestijnenlaan
       200B, 3001 Leuven, Belgium\\
       $^4$Department of Astrophysics, Radboud University Nijmegen, PO Box 9010, 6500GL
       Nijmegen, The Netherlands\\
       $^5$Centre for Astrophysics, University of Central Lancashire, Preston, PR1 2HE,
       UK\\
       $^6$Astrophysics Group, Cavendish Laboratory, Cambridge University, Cambridge, CB3 0HE, UK\\}

\date{\today}

\pagerange{\pageref{firstpage}--\pageref{lastpage}} \pubyear{2005}

\begin{document}

\maketitle

\label{firstpage}

\begin{abstract}
The triple system HD\,158926 ($\lambda$\,Sco) has been observed
interferometrically with the Sydney University Stellar Interferometer and the
elements of the wide orbit have been determined. These are significantly more
accurate than the previous elements found spectroscopically. The inclination of
the wide orbit is consistent with the inclination previously found for the orbit
of the close companion. The wide orbit also has low eccentricity, suggesting
that the three stars were formed at the same time.

The brightness ratio between the two B stars was also measured at $\lambda =$
442\,nm and 700\,nm. The brightness ratio and colour index are consistent with the
previous classification of $\lambda$ Sco A as B1.5 and $\lambda$ Sco B as B2.
Evolutionary models show that the two stars lie on the main sequence. Since they
have have the same age and luminosity class (IV) the mass-luminosity relation can be
used to determine the mass ratio of the two stars: $M_B/M_A = 0.76\pm0.04$.

The spectroscopic data have been reanalyzed using the interferometric values for
$P$, $T$, $e$ and $\omega$, leading to revised values for $a_1\sin i$ and the mass
function.  The individual masses can be found from the mass ratio, the mass
function, spectrum synthesis and the requirement that the age of both components
must be the same:  $M_A = 10.4\pm 1.3\,M_\odot$ and $M_B = 8.1\pm 1.0\,M_\odot$.

The masses, angular semimajor axis and the period of the system can be used to
determine the dynamical parallax. We find the distance to $\lambda$\,Sco to be
$112\pm 5$\,pc, which is approximately a factor of two closer than the HIPPARCOS
value of $216\pm42$\,pc.
\end{abstract}


\section[]{Introduction}\label{sec:1}

The bright southern star $\lambda$\,Scorpii (HD 158926, HR 6527, $\alpha_{2000} =
17^{\rm h}33^{\rm m}36\fs52$, $\delta_{2000} = -37^\circ06\arcmin13\farcs8$, $V =
1.62$) is classified as a B2IV+B single-lined spectroscopic binary system in the
Bright Star Catalogue \cite{hoffleit82}, but it is in fact a triple system. Two
recent papers by Uytterhoeven and colleagues \cite{U1,U2} present a detailed
spectroscopic study of the system based on fourteen years of data. We shall refer to
these as Papers I \& II. These papers also include an extensive review of the
literature on $\lambda$\,Sco and here we present only a brief synopsis.

The system comprises two B stars of similar mass that orbit each other with a
period of approximately 1000 days (the `wide' system). The primary is a
$\beta$\,Cephei type variable with a low mass companion that orbits the primary
with a period of 6 days (the `close' system).  The spectroscopic elements for
the two orbits are given in Paper~I. The close binary system is eclipsing and in
Paper~I it is shown that this constrains the inclination to the range $70^\circ
< i_{\rm close} < 90^\circ$.  An independent argument in Paper~II based on
frequency analysis also gives the same result.

Paper~II notes that the wide orbit cannot be accurately determined from the
spectroscopy.  We report here interferometric observations made with the Sydney
University Stellar Interferometer (SUSI) on the wide system that results in a
much improved orbit.

In Section \ref{sec:susi} we present the SUSI observations and the new
interferometric orbit.

In Section \ref{sec:combined} the interferometric elements are used to determine a
revised spectroscopic orbit. Interferometry also provides the brightness ratio
between the distant component and the primary.  The mass-luminosity relation can be
used to find the mass ratio, and using the spectroscopically determined mass
function, the individual masses can be calculated. Having determined the masses the
dynamical parallax for $\lambda$\,Sco can be calculated. It differs significantly
from the HIPPARCOS parallax.

Following Papers I \& II we use `close' and `wide' to distinguish the orbital
elements associated with the 6-day and 1000-day periods seen in the spectrum of
the primary. When no subscript is used the wide system is implied.

In the case of the interferometric data we follow the scheme proposed in the
Washington Multiplicity Catalog (WMC) \cite{hartkopf03}.  The primary will be
denoted by `A', the close companion by `a' and the wide component by `B'.  The
notation $a_{Aa}$, for example, denotes the semimajor axis of the orbit of a
relative to the primary while $a_{Aa,B}$ is the semimajor axis of B relative to the
centre of light of the close pair.

\section{The SUSI observations and the interferometric orbit}\label{sec:susi}

\subsection{Observations with the `blue' beam-combining optics}\label{sec:blue}

Observations prior to 2004 were made at the blue wavelength of 442\,nm using a
bandwidth of 4\,nm.  The optical layout was essentially the same as the one
described in Davis et al.\ \shortcite{davis1999a}.  The data were reduced using
the same procedure that was employed for the analysis of the binary system
$\beta$\,Cen \cite{davis2005a}.

Baselines of 5, 10 and 20 metres were used. For these baselines the two B stars
are essentially unresolved by SUSI. Because of atmospheric turbulence the fringe
phase carries no useful information, and SUSI, like most optical
interferometers, measures a quantity proportional to the squared modulus of the
complex degree of coherence $\gamma$:
\begin{equation}\label{eq:gamma}
    |\gamma|^2 =\frac{1 + \beta^2 + 2\beta\cos(2\upi{\bf b}\!\cdot\!\brho/\lambda)}{(1+\beta)^2}
\end{equation}
where $\bf b$ is the projected baseline, $\brho$ is the vector separation between
the two components of the binary as seen from the Earth and $\beta = I_B/I_A \leq1$
is the brightness ratio.

The SUSI data taken with the blue beam-combining system are `uncalibrated';
i.e., no unresolved reference stars were observed to take into account the
reduction in the observed fringe signal due to atmospheric and instrumental
effects. The seeing conditions vary during the night, and experience has shown
that modelling this by a simple linear time variation is usually adequate. We
therefore assume that the observed signal will have the form
\begin{equation}\label{eq:bluvis}
    |V|^2 = (C_a - C_bt)|\gamma|^2
\end{equation}
where $t$ is the hour angle of $\lambda$\,Sco. The two empirical factors $C_a$
and $C_b$ are free parameters in the model fitting procedure. The sign in
equation (\ref{eq:bluvis}) reflects the fact that in general the seeing
deteriorates during the night.

It should be noted that the cosine term in equation (\ref{eq:gamma}) depends on
the position angle $\theta$ and the separation $\rho$. When the raw $|V|^2$ data
are fitted using equation (\ref{eq:bluvis}) the resultant estimates for $\rho$
and $\theta$ are not significantly affected by the fact that the data are
uncalibrated.  The estimate for the brightness ratio, on the other hand, may be
subject to systematic errors.

The presence of a cosine in equation (\ref{eq:gamma}) means that there is a
$180^\circ$ ambiguity in the position angle (a two-aperture interferometer
cannot distinguish between $\brho$ and $-\brho$).  This has no impact on the
interpretation of the results presented here.

Equation (\ref{eq:gamma}) does not take into account the presence of the close
binary orbiting the primary. However, Paper~I notes that the contribution of
H$_\alpha$ emission from weak T Tauri stars is at least 2 orders of magnitude
fainter than the primary.  Modulation due to the faint companion will be at most 1\%
and generally much smaller than this.  This is less than the measurement noise and
consequently the effect of the T Tauri component on the interferometric data will be
negligible.

\begin{figure}
 \centering
 \includegraphics[width=84mm,clip]{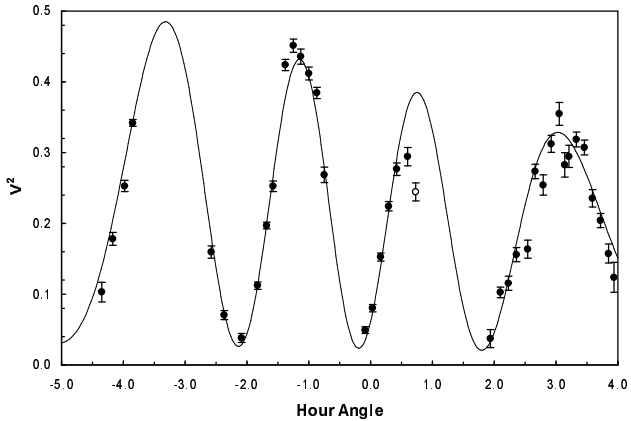}
 \caption{A typical set of data from one night (27 June 2000). Each point
 represents a measurement of $|V|^2$.  The data are fitted using equation
 (\ref{eq:bluvis}) and the best fit is shown. The error bars denote the
 formal errors of the data points. The actual uncertainties are estimated
 to be 2.8 times larger. The point shown as an open circle was omitted
 from the final fit.}
 \label{fig:example}
\end{figure}

The binary star measures were estimated from the $|V|^2$ data using essentially
the same procedure described by Davis et al. \shortcite{davis2005a}. The data
from each night were fitted using equation (\ref{eq:bluvis}) and the data for a
typical night are shown in Fig.~\ref{fig:example}.  The formal error for each
data point in general underestimates the actual scatter because the seeing
varies throughout the night. Having found the best fit to the data the formal
errors are rescaled to make $\chi^2 = \nu$, where $\chi^2$ is the weighted sum
of the squared residuals and $\nu$ is the number of degrees of freedom. In
effect, we assume that the actual uncertainties obey Gaussian statistics and all
the data points are similarly affected by seeing (in the example shown in
Fig.~\ref{fig:example} the scaling factor was 2.8). Using the rescaled weighting
factors Monte Carlo simulation is then used to determine the uncertainties in
the fitting parameters.

The assumption that all the points are equally affected by seeing is not
strictly true, but if the seeing does not change dramatically during the night
this assumption is, on average, realistic.  The seeing is continually measured
and logged, and points that are badly affected by seeing are excluded from the
final fit.

A completely independent analysis was performed using simulated annealing
\cite{kirkpatrick83}. Simulated annealing is more likely to converge on the true
minimum of $\chi^2$ rather than on a local minimum, unlike steepest descent methods.
The simulated annealing results did not differ significantly from those obtained by
our standard method and we are confident that the values of $\rho$, $\theta$ and
$\beta$ and their uncertainties that are given in Table \ref{table:susidata} are the
best fit to each night's data. The separation $\rho$ and related quantities are all
expressed in milliarcseconds (mas).

\begin{table*}
 \begin{minipage}{160mm}
  \caption{The binary star measures determined with SUSI. The date, Julian date,
  the baseline used and the wavelength are tabulated along with the measured
  values of $\rho$, $\theta$, $\beta$ and their uncertainties.  The weighting
  factor $W$ was used when fitting the orbit and is a measure of the accuracy
  of each point.  The final column gives the residual between the observed data
  and the position calculated from the best fitting model. Using the WMC
  nomenclature, all these observations are classified as Aa,B.}

  \label{table:susidata}
  \begin{tabular}{@{}r r r r r r r r  r r @{}l r @{\,} r @{\,} c }
  \hline
  \multicolumn{1}{c}{Date} &
  \multicolumn{1}{c}{JD} &
  \multicolumn{1}{c}{$B$}&
  \multicolumn{1}{c}{$\lambda$} &
  \multicolumn{1}{c}{$\Delta\lambda$} &
  \multicolumn{1}{c}{$\rho$} &
  \multicolumn{1}{c}{$\sigma_\rho$} &
  \multicolumn{1}{c}{$\theta$} &
  \multicolumn{1}{c}{$\sigma_\theta$} &
  \multicolumn{1}{c}{$\beta$}& &
  \multicolumn{1}{c}{$\sigma_\beta$}&
  \multicolumn{1}{@{\,}c@{\,}}{$W$}&
  \multicolumn{1}{c}{$|O - C|$}\\
  \multicolumn{1}{c}{(UT)}&
  &
  \multicolumn{1}{c}{(m)}&
  \multicolumn{1}{c}{ (nm)} &
  \multicolumn{1}{c}{(nm)} &
  \multicolumn{1}{c}{(mas)} &
  \multicolumn{1}{c}{(mas)} &
  &
  &
  &
  &
  &
  \multicolumn{1}{@{\,}c@{\,}}{(mas$^{-1}$)}
  &\multicolumn{1}{c}{(mas)}\\
 \hline
1999-05-15  &   2451314.1   &   10  &   442 &   4   &   45.86   &   0.19    &   265 \fdg    92  &   0   \fdg    06  &   0.532   &&   0.008   &   5.0 &   0.51    \\
1999-05-17  &   2451316.1   &   5   &   442 &   4   &   45.66   &   0.27    &   266 \fdg    37  &   0   \fdg    08  &   0.596   &&   0.009   &   3.6 &   0.44    \\
1999-06-04  &   2451334.1   &   5   &   442 &   4   &   46.12   &   0.18    &   267 \fdg    95  &   0   \fdg    07  &   0.649   &&   0.009   &   5.2 &   0.73    \\
1999-06-22  &   2451352.1   &   5   &   442 &   4   &   46.79   &   0.97    &   268 \fdg    93  &   0   \fdg    27  &   0.540   &&   0.013   &   1.0 &   0.50    \\
1999-06-23  &   2451353.1   &   5   &   442 &   4   &   46.25   &   0.83    &   268 \fdg    50  &   0   \fdg    26  &   0.580   &&   0.023   &   1.2 &   1.11    \\
1999-08-12  &   2451403.1   &   5   &   442 &   4   &   44.53   &   1.82    &   273 \fdg    09  &   0   \fdg    43  &   0.482   &&   0.020   &   0.5 &   1.40    \\
2000-05-11  &   2451676.1   &   10  &   442 &   4   &   25.73   &   0.21    &   72  \fdg    46  &   0   \fdg    19  &   0.553   &&   0.015   &   4.5 &   0.34    \\
2000-05-12  &   2451677.1   &   10  &   442 &   4   &   26.08   &   0.14    &   72  \fdg    73  &   0   \fdg    10  &   0.596   &&   0.006   &   6.6 &   0.29    \\
2000-05-15  &   2451680.1   &   10  &   442 &   4   &   26.96   &   0.16    &   73  \fdg    95  &   0   \fdg    11  &   0.590   &&   0.019   &   5.9 &   0.16    \\
2000-05-18  &   2451683.1   &   10  &   442 &   4   &   28.33   &   0.11    &   74  \fdg    74  &   0   \fdg    09  &   0.609   &&   0.003   &   8.4 &   0.52    \\
2000-05-27  &   2451692.1   &   10  &   442 &   4   &   30.23   &   0.23    &   76  \fdg    49  &   0   \fdg    19  &   0.537   &&   0.010   &   4.0 &   0.18    \\
2000-06-02  &   2451698.1   &   10  &   442 &   4   &   31.43   &   0.14    &   77  \fdg    56  &   0   \fdg    13  &   0.528   &&   0.005   &   6.4 &   0.06    \\
2000-06-03  &   2451699.1   &   10  &   442 &   4   &   31.45   &   0.10    &   77  \fdg    57  &   0   \fdg    07  &   0.600   &&   0.002   &   9.0 &   0.29    \\
2000-06-04  &   2451700.1   &   10  &   442 &   4   &   31.25   &   0.10    &   77  \fdg    72  &   0   \fdg    05  &   0.575   &&   0.005   &   10.0    &   0.72    \\
2000-06-07  &   2451703.1   &   10  &   442 &   4   &   32.41   &   0.13    &   78  \fdg    35  &   0   \fdg    06  &   0.529   &&   0.008   &   7.3 &   0.26    \\
2000-06-11  &   2451707.1   &   10  &   442 &   4   &   33.45   &   0.28    &   78  \fdg    83  &   0   \fdg    14  &   0.468   &&   0.008   &   3.4 &   0.21    \\
2000-06-27  &   2451723.1   &   10  &   442 &   4   &   36.94   &   0.14    &   81  \fdg    39  &   0   \fdg    07  &   0.613   &&   0.014   &   6.7 &   0.10    \\
2000-06-30  &   2451726.1   &   10  &   442 &   4   &   37.44   &   0.24    &   81  \fdg    83  &   0   \fdg    14  &   0.455   &&   0.010   &   3.9 &   0.20    \\
2000-07-01  &   2451727.1   &   5   &   442 &   4   &   40.00   &   0.40    &   82  \fdg    08  &   0   \fdg    18  &   0.588   &&   0.014   &   2.4 &   2.16    \\
2000-07-22  &   2451748.1   &   5   &   442 &   4   &   43.54   &   0.21    &   84  \fdg    10  &   0   \fdg    09  &   0.560   &&   0.009   &   4.4 &   1.83    \\
2000-08-02  &   2451759.1   &   5   &   442 &   4   &   44.30   &   0.26    &   85  \fdg    30  &   0   \fdg    16  &   0.543   &&   0.012   &   3.5 &   0.83    \\
2001-06-22  &   2452083.1   &   10  &   442 &   4   &   18.80   &   0.17    &   129 \fdg    16  &   0   \fdg    43  &   0.534   &&   0.012   &   4.5 &   1.14    \\
2001-08-10  &   2452132.1   &   10  &   442 &   4   &   12.97   &   0.16    &   163 \fdg    25  &   1   \fdg    61  &   0.563   &&   0.018   &   2.5 &   0.62    \\
2001-08-19  &   2452141.1   &   20  &   442 &   4   &   12.58   &   0.04    &   174 \fdg    46  &   0   \fdg    43  &   0.476   &&   0.009   &   9.6 &   0.54    \\
2004-04-15  &   2453111.2   &   20  &   700 &   80  &   24.05   &   0.20    &   118 \fdg    92  &   0   \fdg    24  &   0.544   &&   0.001   &   4.5 &   0.69    \\
2004-04-20  &   2453116.2   &   20  &   700 &   80  &   23.38   &   0.24    &   119 \fdg    37  &   0   \fdg    51  &   0.518&$^a$   &   0.028   &   3.1 &   0.34    \\
2004-06-24  &   2453181.0   &   20  &   700 &   80  &   12.77   &   0.03    &   160 \fdg    40  &   0   \fdg    30  &   0.488   &&   0.006   &   13.3    &   0.26    \\
2005-04-17  &   2453478.2   &   20  &   700 &   80  &   47.42   &   0.20    &   270 \fdg    36  &   0   \fdg    05  &   0.516   &&   0.014   &   4.8 &   0.18    \\
2005-07-18  &   2453570.0   &   15  &   700 &   80  &   37.83   &   0.37    &   279 \fdg    45  &   0   \fdg    10  &   0.635&$^a$    &   0.019   &   2.6 &   0.17    \\
2005-07-21  &   2453573.0   &   15  &   700 &   80  &   37.18   &   0.23    &   279 \fdg    76  &   0   \fdg    05  &   0.536&$^a$    &   0.013   &   4.2 &   0.24    \\
2005-08-17  &   2453600.0   &   15  &   700 &   80  &   32.23   &   0.28    &   283 \fdg    68  &   0   \fdg    07  &   0.543&$^a$    &   0.008   &   3.5 &   0.45    \\

\hline
\end{tabular}

\medskip
$^a$ $V^2_{\rm max} < 1$ for these sets of data. The other observations at 700\,nm
were fitted using $|V|^2 =1$. This is discussed in Section \ref{sec:red}.
\end{minipage}
\end{table*}

\subsection{Observations with the `red' beam-combining optics}\label{sec:red}

The more recent SUSI data have been obtained with the red beam-combining optics.
This system has been described in Davis et al. \shortcite{davis2005b}.  The
observations were made using baselines in the range 10--20 metres.  Again the
individual B stars are unresolved at these baselines and the effect of the
six-day companion on the data will be negligible.

The red observations were made at $\lambda = 700$\,nm using a bandwidth of
80\,nm.  With such a wide bandwidth the effects of ``bandwidth smearing'' cannot
be ignored.  Tango \& Davis \shortcite{tango2002} have discussed bandwidth
smearing in the context of single stars and it is straightforward to generalize
their results to arbitrary systems. In particular, if $\nu = c/\lambda$ is the
optical frequency, $S_\nu$ is the spectral response of the interferometer and
\begin{equation}
    N_\nu = \int\!\!\!\int\!d^2\mathbf{x} I_\nu(\mathbf{x})
\end{equation}
is the integrated monochromatic flux from the source (the integral is taken over
the plane of the sky), the broad band complex degree of coherence will be
\begin{equation}\label{eq:vCZnu}
\overline \gamma = \frac{\int_0^\infty d\nu S_\nu N_\nu \gamma(\nu)} {\int_0^\infty
d\nu S_\nu N_\nu}
\end{equation}
where $\gamma(\nu)$ is the quasimonochromatic complex degree of coherence
defined by
\begin{equation}\label{eq:vCZ}
\gamma(\nu) = N_\nu^{-1}\int\!\!\!\int\!d^2\mathbf{x}\exp\{-2\upi  i
\mathbf{b\!\cdot\! x}c^{-1}\nu\}I_\nu(\mathbf{x})
\end{equation}
The complex coherence $\gamma(\nu)$ will vary with frequency (a) because of the
frequency dependence of the Fourier kernel and (b) because $I_\nu({\mathbf
x})/N_\nu$ is also a function of the frequency. In the case of a binary system
with unresolved components this latter effect will be unimportant.

We assume that the bandwidth is rectangular, having a width $\delta\nu$ centred
on $\nu_0$.  Let $\lambda_0 = c/\nu_0$ and make the usual approximation that
$\delta\nu \approx c\delta\lambda/\lambda_0^2$.  For convenience define $\psi =
2\upi{\bf b}\!\cdot\!\brho/\lambda_0$.  The equivalent of equation
(\ref{eq:gamma}) including bandwidth smearing becomes:
\begin{equation}\label{eq:redvis}
    |\gamma|^2=
    \!\frac{1 + \beta^2{\rm\;sinc}^2(\psi\delta\lambda/2\lambda_0) + 2\beta{\rm \;sinc}
    (\psi\delta\lambda/2\lambda_0)\cos\psi}{(1+\beta)^2}
\end{equation}

In the case of $\lambda$\,Sco the effect of bandwidth smearing can reduce
$|\gamma|^2$ by several percent.  The effect is time-dependent, and is most
significant when $\bf b$ and $\brho$ are parallel.

The red table observations were first processed using the ``data pipeline''
described in Davis et al. \shortcite{davis2005b}.  The output is a set of
calibrated measures of $|V|^2$.  The visibility measures for each night were
fitted using the same procedure adopted for the `blue' observations.  The only
difference is that the red data are fitted using equation (\ref{eq:redvis})
rather than equation (\ref{eq:bluvis}).  The most noticeable difference between
the blue and red data sets is that the errors in the red data points required
very much less rescaling, confirming the fact that the use of calibrators
significantly reduces the effects of seeing on the data.

If the data are properly calibrated the maximum value of $|V|^2$, $|V|^2_{\rm
max}$, should be 1.  On several nights, however, it was found that allowing
$|V|^2_{\rm max}$ to be a free parameter gave significantly better fits, with
$|V|^2_{\rm max} < 1$. The reduction cannot be explained by the partial
resolution of the B stars or the modulation caused by the 6-day close companion.
We believe that the effect is most likely instrumental but there are as yet not
enough data to establish the exact cause. This effect has a negligible effect on
the values of $\rho$ and $\theta$, but may affect $\beta$.

The results from the observations made with the red beam-combining system are
given in Table \ref{table:susidata}.

\subsection{The interferometric orbit}
\begin{figure*}
 \centering
 \includegraphics[width=177mm,clip]{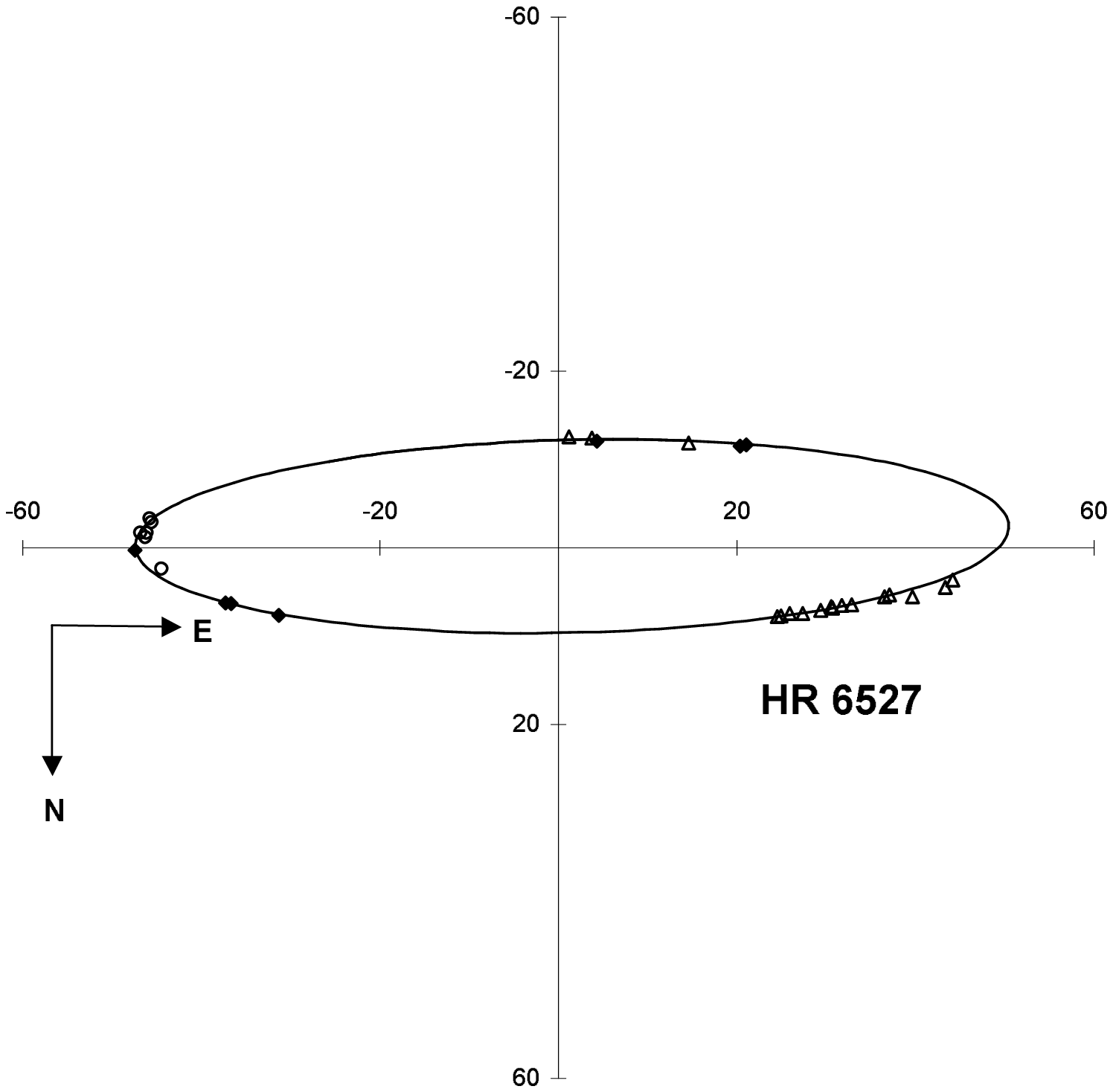}
 \caption{The orbit of $\lambda$\,Sco. The scale is in millarcsec and the
 orientation conforms to standard practice \protect\cite{heintz78}.  Different
 symbols are used to denote successive orbits, referred to the epoch of
 periastron $T$. The circles indicate data taken in 1999.  The triangles are the
 data from 2000 and 2001. The diamonds denote points on the next orbit during
 2004 and 2005. Open symbols are used for observations at 442\,nm; filled
 symbols denote 700\,nm observations.}
 \label{fig:orbit}
\end{figure*}

The orbital elements were calculated, again using the same procedure that was
employed for $\beta$\,Cen \cite{davis2005a}.  Each measure in Table
\ref{table:susidata} was given a weight according to the formula: $W =
(\sigma_\rho^2 + \rho^2\sigma_\theta^2)^{-1/2}$.  The results are given in Table
\ref{table:elements}, and the `$|O-C|$' residuals are tabulated in Table
\ref{table:susidata}.

The uncertainties in the orbital elements were estimated using Monte Carlo
simulation using the best fit orbital elements and the weighting factors in Table
\ref{table:susidata} to generate the simulated data sets. The resultant
uncertainties are also listed in Table \ref{table:elements}.

The period $P$, epoch of periastron $T$ and eccentricity $e$ differ
significantly from those given in Paper~I (see Table \ref{table:spec}), and we
return to this in Section \ref{sec:combined}.

The computed orbit and the observed measures are shown in
Fig.~\ref{fig:orbit}. This figure also confirms the fact that any effects caused
by the close companion must be smaller than the fitting errors.

\begin{table}
\begin{center}
 \caption{The orbital elements of $\lambda$\,Sco determined from the SUSI
 observations. All the elements refer to the Aa,B orbit. }
 \label{table:elements}
 \begin{tabular}{ c @{} r@{$\;\pm\;$}l l@{}}
 \hline
 Element & \multicolumn{3}{c}{Value, uncertainty and units}\\
 \hline
 $P$&1052.8&1.2& d\\
 $T$&51\,562.3&2.8& MJD\\
 $e$&0.121&0.005& \\
 $a\arcsec$&49.3&0.3&mas\\
 $i$&77$\fdg$2&0$\fdg$2& \\
 $\omega$&74$\fdg$8&0$\fdg$9&See note. \\
 $\Omega$&271$\fdg$30&0$\fdg$15& \\
 $(a\arcsec)^3/P^2$ &\multicolumn{2}{r}{$(108\pm2)\times10^{-12}$}&asec$^3$d$^{-2}$\\
  \hline
 \end{tabular}
 \end{center}
 \medskip
 {\it Note:~~}This is the argument of periastron for the orbit of the  B
 component relative to the primary.  The corresponding quantity in Table
 \ref{table:spec} refers to the orbit of the primary around the system
 barycentre. The two angles differ by 180$^\circ$.
\end{table}

Our value of the inclination, $i_{Aa,B} = 77\fdg2\pm0\fdg2$, is completely
consistent with the range $70^\circ\leq i_{\rm close} \leq 90^\circ$ given in
Papers I \& II and, together with the relatively low eccentricity of the
interferometric orbit, supports the hypothesis that the $B$ component did not
join the system as a result of a tidal capture event.

\subsection{The brightness ratio and the mass ratio}

\begin{figure}
 \centering
 \includegraphics[width=84mm,clip]{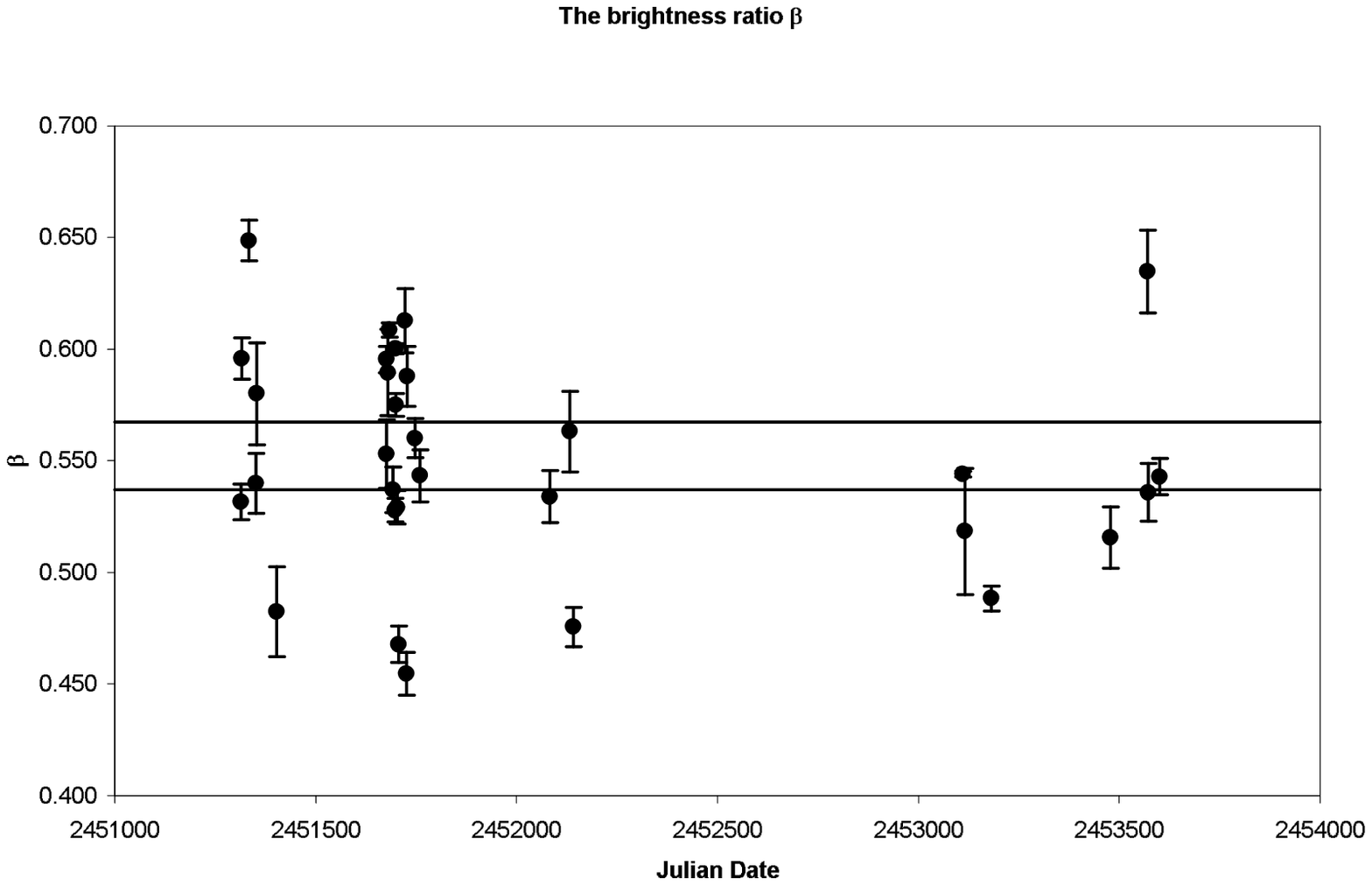}
 \caption{The brightness ratios, taken from Table~\ref{table:susidata}. The two
 lines denote the range in $\beta$ that is expected due to the intrinsic
 variability of the system. The data prior to JD 2\,452\,500 were measured at
 $\lambda = 442$\,nm.  The more recent data were measured at 700\,nm.}
 \label{fig:ratio}
\end{figure}

For the reasons discussed in Section \ref{sec:blue} the brightness ratio $\beta$
is subject to systematic errors associated with the way the $|V|^2$ data are
normalized. As well, the system is a $\beta$ Cephei variable with a photometric
amplitude in $v$ of approximately 0.023 magnitudes and a period of 5\fh1.
\cite{shobbrook72,shobbrook75}.

Shobbrook \& Lomb \shortcite{shobbrook75} also observed periodic variations in
colour.  The amplitudes of the variation were $(b - v) = 0.002\pm0.001$ and $(u
- b) = 0.0098\pm0.0012$.

The measured brightness ratios and their uncertainties are plotted in
Fig.~\ref{fig:ratio}. The data fall into two groups: the earlier set of points
were observed using $\lambda = 442$\,nm while the more recent set was obtained
using $\lambda = 700$\,nm. The two lines represent the approximate spread in
$\beta$ that would be expected from the intrinsic variability ($\sim\!\pm0.012$
mag).  The error bars indicate the statistical uncertainties for each
observation, but in the present context they are misleading, since there are
significant night-to-night variations due to the intrinsic variability of
$\lambda$\,Sco as well as the instrumental effects (calibration errors)
previously discussed. In our view the average and standard error of the blue and
red data sets represent the best estimates of the brightness ratios.

The means and standard deviations for the brightness ratio for each set, and the
corresponding magnitude differences, are
\begin{eqnarray}
\beta_{442} = 0.55\pm0.05&~~~~&\Delta m_{442} = 0.65\pm0.10\label{eq:betaB}\\
\beta_{700} = 0.54\pm0.05&~~~~&\Delta m_{700} = 0.67\pm0.10\label{eq:betaR}
\end{eqnarray}
The difference in the colour index $(m_{442}-m_{700})$ between $\lambda$ Sco A and B
is $-0.02\pm0.14$. Both the magnitude and colour index differences are consistent
with the relative spectral classification of $\lambda$\,Sco A as B1.5 IV and
$\lambda$ Sco B as B2 V given in Paper~I.

Although the A and B components have apparently different luminosities, due to the
complexity of this system and the fast rotation there is an uncertainty of up to 0.2
dex in $\log g$, making it difficult to establish the luminosity class with
certainty. However, the inclination of the long period orbit suggests that the three
stars formed together and $\lambda$ Sco A and B are thus both on the main sequence.
The brightness ratio then implies that the effective temperatures can be fixed at
25\,000\,K and 21\,000\,K. The contribution of the T Tauri star to the luminosity of
the primary is negligible and the empirical mass-luminosity relation $\log (L_B/L_A)
= (3.51\pm0.14)\log (M_B/M_A)$ \cite{Griffiths88} can be used to find the mass ratio
for $\lambda$\,Sco A and B. The ratio of the luminosities has been obtained from the
V magnitude difference, taken to be 0.66$\pm$0.10 by interpolation between $\Delta
m_{442}$ and $\Delta m_{700}$, and the difference in bolometric correction (BC).
The latter has been obtained from interpolation in the tables by Flower
\shortcite{flower96} for main-sequence B stars.  Assuming an uncertainty of
$\pm$1000\,K in the effective temperatures as obtained in Paper~I, we arrive at
(BC$_{A}$ - BC$_{B}$) = -0.40$\pm$0.10.  Thus the mass ratio for $\lambda$\,Sco A
and B is:
\begin{equation}\label{eq:ratio}
    \frac{M_B}{M_A} = 0.76\pm0.04
\end{equation}
This is consistent with the value of $M_B/M_A = 0.84\pm0.12$ estimated from the
spectrum synthesis analysis presented in Paper~I.

\section{The combined interferometric and spectroscopic orbit}
\label{sec:combined}

\begin{figure}
 \centering
 \includegraphics[width=84mm,clip]{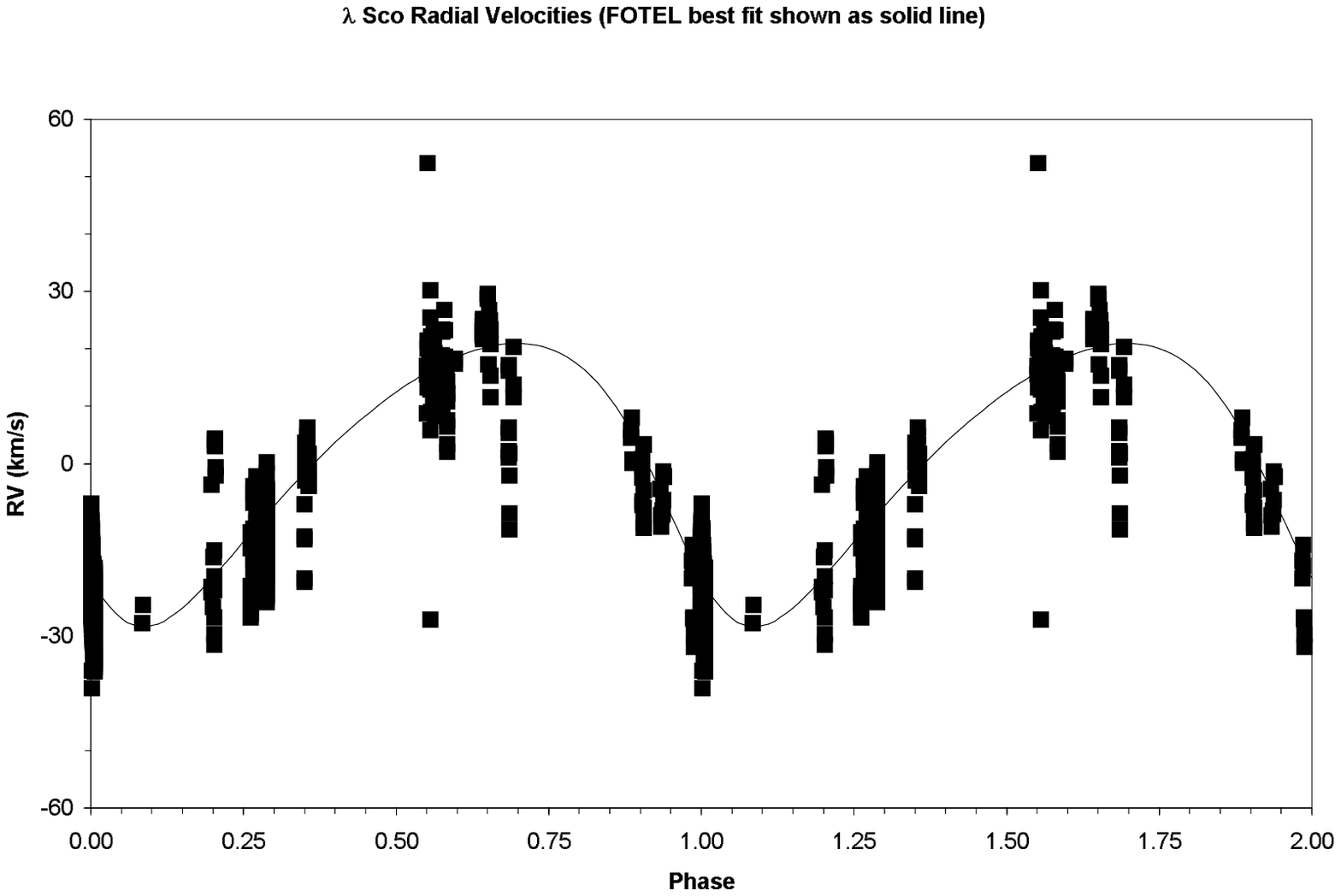}

 \medskip
 \includegraphics[width=84mm,clip]{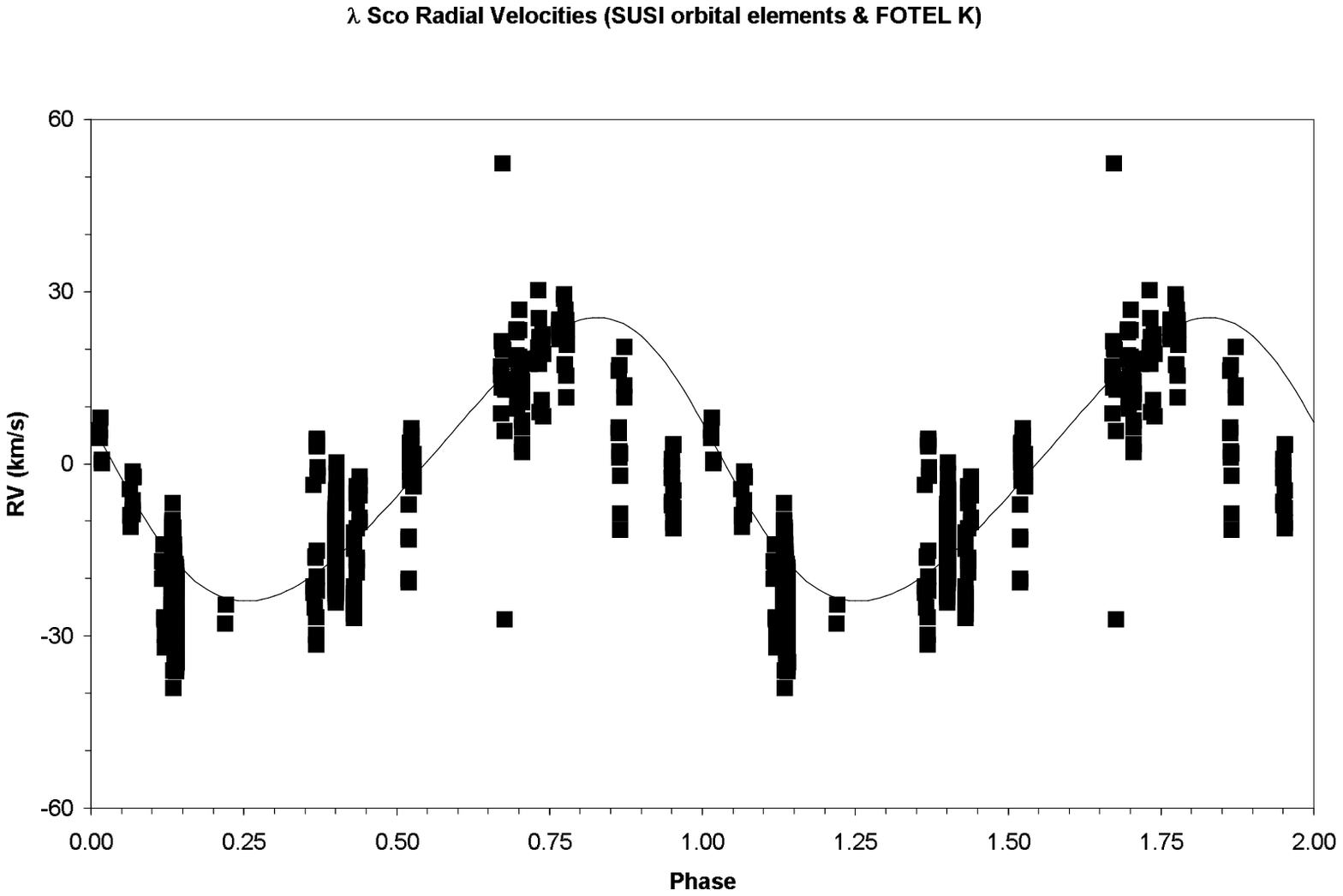}

 \caption{The radial velocity data for the `wide' system in $\lambda$\,Sco. The
 RV due to the `close' system has been removed. In the upper panel the data are
 plotted against the FOTEL phase and the RV curve is calculated from the FOTEL
 elements in Table~(\ref{table:spec}).  The same RV data are plotted in the lower panel
 against the
 SUSI phase, with the RV curve calculated using the elements given in the final
 column of Table~(\ref{table:spec}).}
 \label{fig:rv-fit}
\end{figure}

The spectroscopic orbital elements were determined in Paper~I using two
different codes, FOTEL and VCURVE. The upper panel of Fig.~\ref{fig:rv-fit}
plots the RV data against the FOTEL phase.  The solid line is the calculated RV
using the FOTEL elements (this panel is the same as the lower panel in Fig.~2 of
Paper~I).

The interferometric values for $P$, $T$, $e$ and $\omega$ are much more accurate
than the spectroscopically determined ones, and the original RV data were
refitted using these values, giving $K_{\rm wide} =
22.8\pm0.4$\,km$\cdot$s$^{-1}$.  Table~\ref{table:spec} tabulates the elements
given in Paper~I and our revised values, and the RV amplitude calculated using
these elements is shown in the lower panel of Fig.~\ref{fig:rv-fit}.

\begin{table}
 \centering
 \caption{The second column lists the elements found in Paper~I using the FOTEL
 code. The third column tabulates the elements used in the final fit: $P$, $T$,
 $e$ and $\omega$ are based on the SUSI; $K$ has been recomputed using these
 values.  The uncertainties are the formal uncertainties of the fit and do not
 include systematic effects.}
 \label{table:spec}
 \begin{tabular}{  l   r@{$\,\pm\,$}l @{$\;$} r@{$\,\pm\,$}l}
 \hline
 Element &\multicolumn{2}{c}{FOTEL}&\multicolumn{2}{c}{FOTEL+SUSI} \\
 \hline
 $P$ (d)& 1082&3& 1052.8&1.2\\
 $K$ (km$\cdot$s$^{-1}$)&24.7&0.4&22.8&0.4\\
 $T$ (MJD)& 51\,731.5&29&51\,562.3&2.8 \\
 $e$&0.23&0.03& 0.121&0.005\\
 $\omega$&311$^\circ$&11$^\circ$&254$\fdg$8&0$\fdg$9\\
 \hline
  \end{tabular}
 \end{table}

Some words of caution on the accuracy of the RV data are in order.  The use of two
types of spectra of very different nature and quality forced the authors of Paper~I
to use different methodology to derive the RV values.  In doing so, they ignored the
fact that the B component contributes to the spectral lines used for the RV
computation, given that the binary is observed to be single-lined. The light, and
hence mass ratio, found from the interferometry is an order of magnitude more
precise than the one in Paper~I and thus allows us to place better constraints on
the individual masses of the  A and B components. The ranges listed in Paper~I must
be revised, as they do not include the systematic effects due to the  B component on
the spectral lines used to determine the amplitude of the RV.

To make the best use of the interferometric results it is important to understand
the uncertainties and possible systematic effects in the RV data.  In general, the
contribution of the  B component to the lines leads to an underestimation of the RV
of the primary. As a consequence, the derived quantities $a\sin i$ and the mass
function are also underestimated. This fact was clearly demonstrated and corrected
for by spectral disentangling in the case of the double-lined binary $\beta$\,Cen
\cite{Ausseloos06}, where it implied a 10\% increase in the individual component
masses. We are unable to apply such a treatment to $\lambda$\,Sco because the
disentangling fails in this single-lined case with three stellar components, of
which at least one is oscillating (see Paper~II). Nevertheless, we can estimate the
systematic errors of the RVs for the two sets of data from which they were derived:
\begin{enumerate}
    \item the high S/N single-order spectra including the Si\,{\sc iii} 4553\AA\
    line taken with CAT/CES;
    \item the low S/N \'echelle spectra taken with Euler/CORALIE.
\end{enumerate}
The resolution for both sets of data is nearly the same (see Paper~I for
details).

From set~(i) we derive that the lines due to the B component are completely situated
within those of the primary (see Paper~I and Waelkens \shortcite{Waelkens90} for
appropriate plots of the lines). Moreover, from the brightness ratio and the
temperature ranges listed in Paper~I we derive that the $\lambda$ Sco B contributes
40\% to the total equivalent width of the Si\,{\sc iii} line used to determine the
RVs. This places strong constraints on the upper limit to $K_{\rm wide}$: the fact
that the B component is not seen in the wings of the high-quality CAT profiles
implies that $K_{\rm wide} < 25.7$\,km$\cdot$s$^{-1}$ (which we deduce from merging
the contributions of the two stars to the Si\,{\sc iii} line with the appropriate EW
ratios and $v\sin i$ estimates of Papers~I and II). This suggests a conservative
range for $K_{\rm wide}$ is $[22.4, 25.7]$\,km$\cdot$s$^{-1}$.
Table~\ref{table:revised} tabulates the system parameters that are affected by the
uncertainties in the radial velocity amplitude.  The accurate value of the mass
ratio given in equation (\ref{eq:ratio}) has been used to calculate the individual
masses from the mass function.

\begin{table}
 \centering
 \caption{ System parameters that depend on the RV amplitude. The lower limit
for $K_{\rm wide}$ is taken from Table \ref{table:spec}.  The upper limit is
based on our best estimates of the effects of the wide component on the RV data
and from the interferometric data.}
 \label{table:revised}
 \begin{tabular}{  l  l l}
 \hline
 \multicolumn{1}{c}{Element}  &
\multicolumn{2}{c}{Range} \\
 \hline
  $K_{\rm wide}$ &[22.4, 25.7] &km$\cdot$s$^{-1}$\\
  $a_1\sin i_{\rm wide}$ &[2.15, 2.47]&AU \\
 $a_1$  & [2.21, 2.53]&AU\\
 $f(M)$ &[1.20, 1.82]&M$_\odot$\\
 $M_A$ &[9.39, 14.25]&M$_\odot$ \\
  $M_B$ &[7.05, 10.69]&M$_\odot$\\
  \hline
 \end{tabular}
 \end{table}

The mass ranges listed in Table~\ref{table:revised}, although compatible with those
given in Paper~I, are very broad. We  tried to constrain them further by taking into
account the results from the spectrum synthesis presented in Paper~I, in which the
contributions from the individual components were properly taken into account to fit
H, He, and Si lines in data set~(ii). We did this by scanning the data base of
standard stellar evolution models computed by Ausseloos et al.\
\shortcite{Ausseloos04}. We selected models from the ZAMS to the TAMS with $X=0.70$
and $Z$ in the range 0.012 to 0.030. Models with the three values 0.0, 0.1 and 0.2
for the core overshooting parameter (expressed in units of the local pressure scale
height) were included. While scanning the database, we required that the two stars
not only satisfy the mass ranges given in Table~\ref{table:revised}, but also that
they have the appropriate luminosity ratio given by the interferometry, and have
$\log T_{\rm eff}$ and $\log g$ in the ranges found from the spectrum synthesis done
in Paper~I. Moreover, we required that their ages be the same to within 0.1\%
(following the interferometric result that the orbital inclinations of the  A and B
components are consistent with each other). In this way, we find the acceptable
range for the mass of the primary $M_A$ to be $[9.0,11.7]\,M_\odot$ and the range
for $M_B$ to be $[7.0,9.1]\,M_\odot$.

All the acceptable model combinations at the lower mass ends of these intervals have
an age below $10^7$~yr corresponding to less than 60\% of the main sequence
lifetime, while those at the upper mass end correspond to an age below $13\times
10^7$~yr which is less than 30\% of the main sequence duration. For the lower mass
end, this is consistent with the suggestion made in Paper~I that $\lambda$ Sco
contains a pre-MS star. Indeed, from Palla \& Stahler \shortcite{palla} we note that
a pre-MS star with a mass of $\sim 1.5\,M_\odot$ remains about $10^7$~yr in its
pre-MS stage when a B-type star formed from the same accretion disk has reached the
ZAMS.  Table~\ref{table:masses} summarizes our conclusions regarding the three
components of $\lambda$\,Sco.

\begin{table}
 \centering
 \caption{ Summary of the stellar parameters of the three compoents of
$\lambda$\,Sco. The mass of the a component is taken from Paper~I; the masses of
$\lambda$\,Sco A and B are based on the work presented here (compare with
Table~4 of Paper~I). }
 \label{table:masses}
 \begin{tabular}{  l  r r r l}
 \hline
 & Mass ($M_\odot$) & $T_{\rm eff}$ (K) & $\log g$ & \\
 \hline
 A & $10.4\pm1.3$& $25\,000\pm1\,000$& $3.8\pm0.1$ & B1.5\,IV\\
 a & $1.8\pm0.2$ & & & pre-MS\\
 B & $8.1\pm1.0$ &$21\,000\pm1\,000$&$4.0\pm0.1$& B2\,IV\\
 \hline
 \end{tabular}
 \end{table}

\subsection{The dynamical parallax}\label{section:parallax}

The dynamical parallax is
\begin{equation}
    \pi_d = \frac{365.25^{2/3}}{(M_{Aa}+M_B)^{1/3}}\left[\frac{(a\arcsec)^3}{P^2}\right]^{1/3}
\end{equation}
where $M_{Aa}$ is the combined mass of the primary (A) and the close companion (a)
and $M_B$ the mass of the B component.  We have used the estimates given in
Table~\ref{table:masses} and the interferometric value of $(a\arcsec)^3/P^2$
(Table~\ref{table:elements}) to find the distance to $\lambda$\,Sco:
\begin{equation}
    D= 1/\pi_d = 112\pm 5~{\rm pc}
\end{equation}

Our distance is smaller than the HIPPARCOS value of $216\pm42$\,pc \cite{HIP1997} by
nearly a factor of two (i.e., 2.5 times the HIPPARCOS standard deviation). It has
been noted \cite{DeZeeuw99} that the motion in binary systems can significantly bias
HIPPARCOS parallaxes and a similar phenomenon has been observed with $\beta$\,Cen
\cite{davis2005a}.

%

The revised parallax may have implications regarding the membership of
$\lambda$\,Sco in the Scorpius-Centaurus-Lupus-Crux complex of OB associations.
Brown \& Verscheuren \shortcite{Brown97} suggested that it was a member of the
Upper Scorpius (US) association, but de Zeeuw et al.\ \shortcite{DeZeeuw99} did
not include $\lambda$\,Sco as a secure member of the US association based on an
analysis of HIPPARCOS positions, parallaxes and proper motions.  The revised
parallax corresponds to the Lower Centaurus Crux (LCC) association, which lies
at a distance of $116\pm2$ pc. The galactic coordinates of $\lambda$\,Sco,
however, still exclude it as a definite member of the association.

\section{Summary}

Observations with SUSI have been used to establish an interferometric orbit for
$\lambda$\,Sco.  The accuracy of the period, epoch of periastron, eccentricity
and argument of periastron are significantly better than those previously found
(Paper~I).  We have combined the interferometric and spectroscopic elements to
provide an orbital solution which is consistent with both the interferometric
and spectroscopic data.  The inclination of the orbit for the wide component is
consistent with that previously found for the close companion, and the
eccentricity of its orbit is quite low.  This is strong evidence that the system
was formed from a common accretion disk rather than through tidal capture.

The brightness ratio allows the mass ratio of the two B stars to be estimated from
the mass-luminosity relation.   The estimated mass ratio of $0.76\pm0.04$ agrees
with the masses determined by the synthetic spectrum analysis presented in Paper~I.
The limiting factor in the determination of the individual masses and the semimajor
axis is clearly the large uncertainties of the RV values. This is due to our
inability to disentangle the contributions of the primary and wide companion to the
lines in the overall spectrum.  While sophisticated disentangling methodology is
available in the literature, e.\,g.\ Hadrava \shortcite{hadrava95}, its application
to single-lined binaries has not yet been achieved to our knowledge. In the case of
$\lambda$\,Sco, the $\beta$\,Cep oscillations are an additional factor that
complicates the disentangling. Nevertheless, we have taken a conservative approach
to estimate the mass ranges from the spectroscopic orbital RV and we have
constrained them further from spectrum synthesis results and from requiring equal
ages and the appropriate brightness ratio from the SUSI data. In this way, we
obtained a relative precision of about 12\% for the masses of both $\lambda$ Sco A
and B.  Masses of B-type stars are not generally known to such a precision.

Accurate mass estimates of pulsating massive stars are important for
asteroseismic investigations of these stars.  Indeed, the oscillation
frequencies of well identified modes in principle allow the fine-tuning of the
physics of the evolutionary models (see Aerts et al.\ \shortcite{aerts03} for an
example). For this to be effective when only a few oscillation modes have been
identified we need an independent high-precision estimate of the effective
temperature and of the mass of the star.  This is generally not available for
single B stars.  The $\lambda$\,Sco Aa system is one of the bright close
binaries with a $\beta$\,Cep component and it is only the second one, after
$\beta$\,Cen, for which a full coverage of the orbit with interferometric data
is available.  The results presented in this paper therefore constitute a
fruitful starting point for seismic modelling of the primary of $\lambda$\, Sco,
particularly now that the WIRE satellite has provided several more oscillation
frequencies compared to those found in Paper~II \cite{brunt05}

The mass estimates, when combined with the period and angular semimajor axis
found from the interferometric orbit, allow the dynamical parallax to be
calculated.  We find the distance to the $\lambda$\,Sco system of $112\pm5$ pc,
which is almost a factor of two smaller than the HIPPARCOS distance of
$216\pm42$ pc. The reason for the large discrepancy is almost certainly due to
the B component since it is of comparable brightness to the primary and the
orbital period is approximately three years.  The fact that many of the B stars
in the Sco-Cen association are multiple systems suggests that the HIPPARCOS
parallaxes for these stars must be used with care.

\section*{Acknowledgments}

This research has been carried out as a part of the Sydney University Stellar
Interferometer (SUSI) project, jointly funded by the University of Sydney and
the Australian Research Council.

MJI acknowledges the support provided by an Australian Postgraduate Award; APJ
acknowledges the support of a Denison Postgraduate Award from the School of
Physics; JRN acknowledges the support of a University of Sydney Postgraduate
Award.

CA and KU are supported by the Fund for Scientific Research of Flanders (FWO)
under grant G.0332.06 and by the Research Council of the University of Leuven
under grant GOA/2003/04.

\bsp

\label{lastpage}

\end{document}